\documentclass{appolb}

\usepackage{graphicx}
\usepackage{bm}
\usepackage{subfigure}
\usepackage{cite}

\makeatletter

\begin{document}

\title{Fluctuations in relativistic heavy-ion collisions from the Glauber models%
\thanks{Presented by WB at {\em Three Days of Strong Interactions},
Wroclaw, 9-11 July 2009, EMMI Workshop and XXVI Max Born Symposium. Research supported in part by the Polish Ministry of Science and Higher
Education, grants N202~034~32/0918 and N202~249235.}
}

\author{Wojciech Broniowski${}^{1,2}$, Maciej Rybczy\'nski${}^{2}$, \L{}ukasz Obara${}^{2}$, Miko\l{}aj Chojnacki${}^{1}$
\address{${}^1$The H. Niewodnicza\'nski Institute of Nuclear Physics, Polish Academy of Sciences, PL-31342 Krak\'ow, Poland\\
${}^2$Institute of Physics, Jan Kochanowski University, PL-25406~Kielce, Poland}
}

\date{19 October 2009}

\maketitle

\begin{abstract}
In the first part of the talk 
we discuss the role of the two-body nucleon-nucleon correlations on signatures of the heavy-ion collisions which are a priori expected to 
be sensitive to these effects. We find that while the fluctuations of the number of produced particles are indeed affected, other quantities 
($v_2$ fluctuations, size fluctuations) are insensitive to the presence of the NN correlations in the nucleon distributions.
In the second part
we show that the fluctuations of the transverse size of the initial source cause, after a suitable hydrodynamic evolution, fluctuations of the 
transverse flow velocity at hadronic freeze-out. This in turn yields the event-by-event fluctuations of the transverse momentum of 
the produced particles, $\langle p_T \rangle$. 
Our results demonstrate that practically
all of the observed event-by-event $\langle p_T \rangle$ fluctuations may be explained this way.
\end{abstract}



\section{Introduction}
This talk consists of two independent parts, both related to novel aspects of the correlations and 
event-by-event fluctuations present in the commonly used 
Glauber treatment of the early phase of the ultra-relativistic heavy-ion collisions.

\section{NN correlations}

The standard way of generating the nucleon distribution  in a nucleus for studies of ultra-relativistic heavy-ion 
collisions is to {\em independently} 
populate the nucleus according to a one-body density in the form of the Woods-Saxon function. Thus, the typically used procedure 
completely neglects the NN {\em correlations} (apart for the poor man's implementation of the  
hard-core repulsion, precluding the centers of the nucleons to be closer 
than a certain distance), known to be crucial in a nuclear system.  
Recently, however, Alvioli, Drescher, and Strikman \cite{Alvioli:2009ab} published nuclear distributions for several nuclei 
which fully account for the central two-body NN
correlations, thus accomplishing the long-awaited task \cite{Baym:1995cz,Heiselberg:2000fk}. 

In this work we use the 
distributions from \cite{Alvioli:2009ab}\footnote{Publicly  available at http://www.phys.psu.edu/$\sim$malvioli/eventgenerator} 
in {\tt GLISSANDO} \cite{Broniowski:2007nz} to investigate within the Glauber treatment 
the role of the NN correlations on several signatures of heavy-ion collisions. We use the wounded-nucleon model \cite{Bialas:1976ed}, 
however we have checked that the results in 
other variants \cite{Broniowski:2007ft} of the Glauber approach are qualitatively similar. Such more detailed studies will be published elsewhere.
We recall that the nucleons from the two nuclei get wounded 
when their centers pass closer to each other than the distance 
$d=\sqrt{\sigma_{\rm NN}/\pi}$, where $\sigma_{\rm NN}$ is the inelastic nucleon-nucleon cross section. For the highest 
SPS, RHIC, and LHC energies it is equal to 32, 42, and 63~mb, respectively.

\begin{figure}[tb]
\begin{center}
\includegraphics[angle=0,width=0.492 \textwidth]{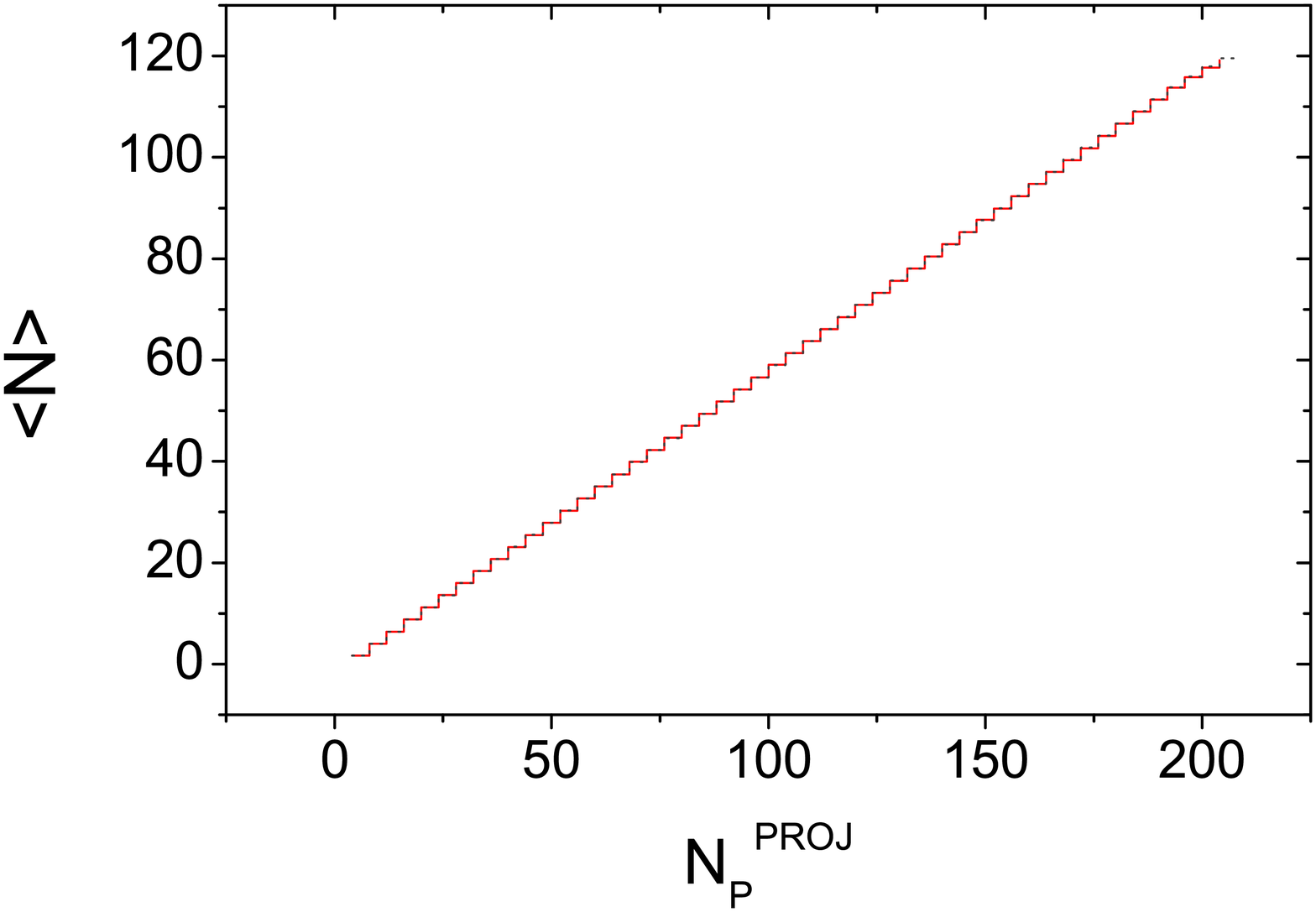}\includegraphics[angle=0,width=0.492 \textwidth]{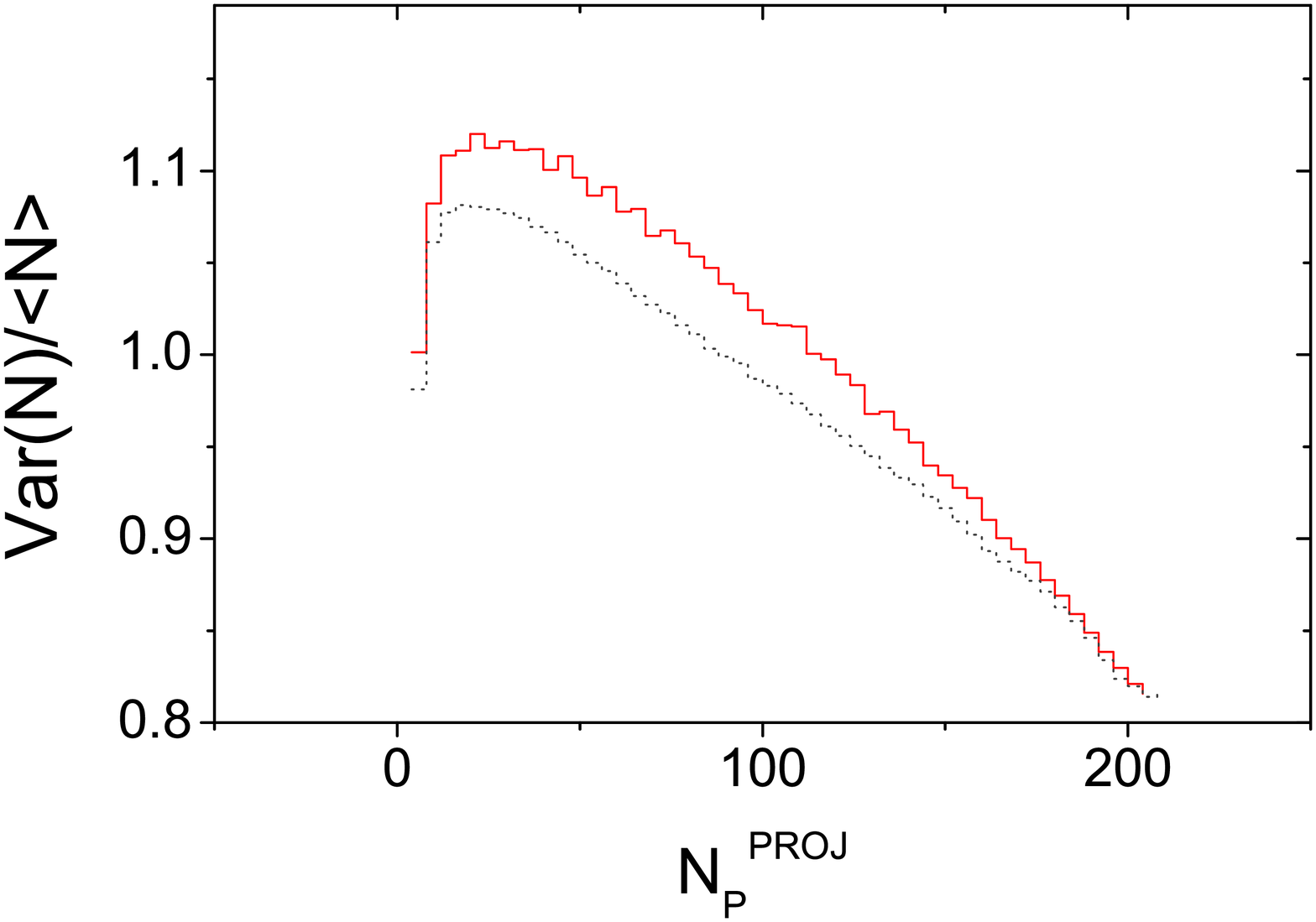}
\\
\includegraphics[angle=0,width=0.492 \textwidth]{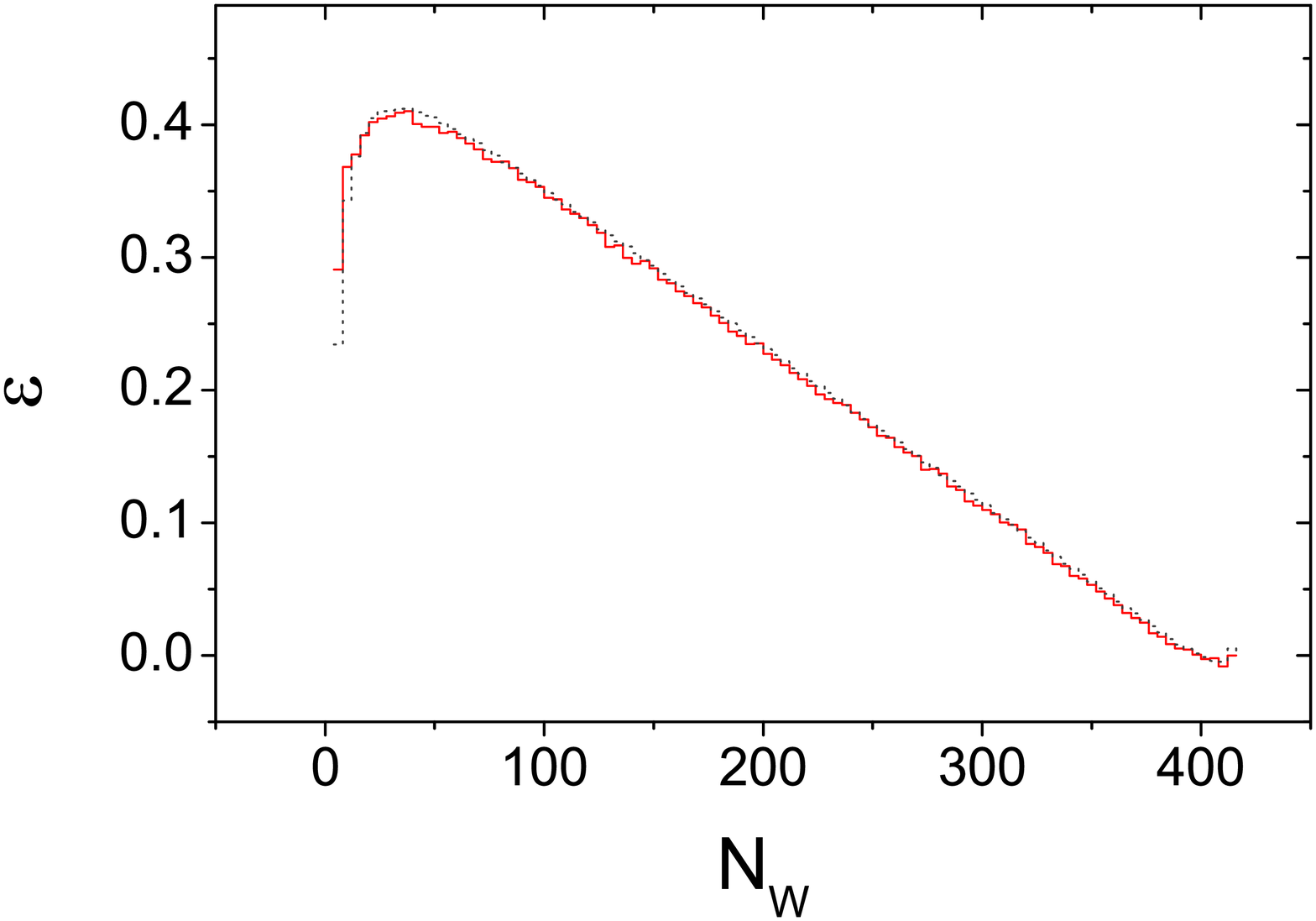}\includegraphics[angle=0,width=0.492 \textwidth]{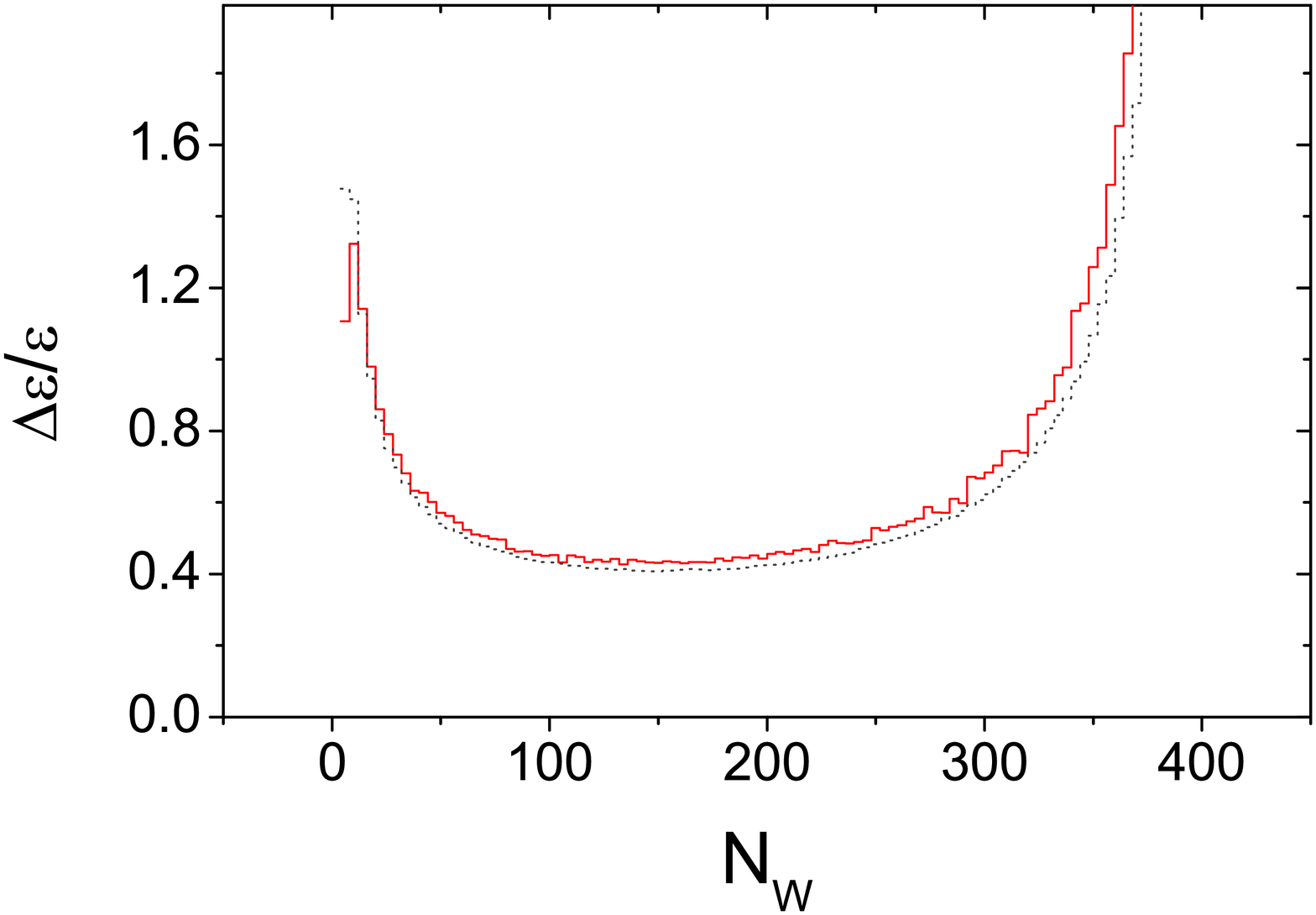}
\\
\includegraphics[angle=0,width=0.492 \textwidth]{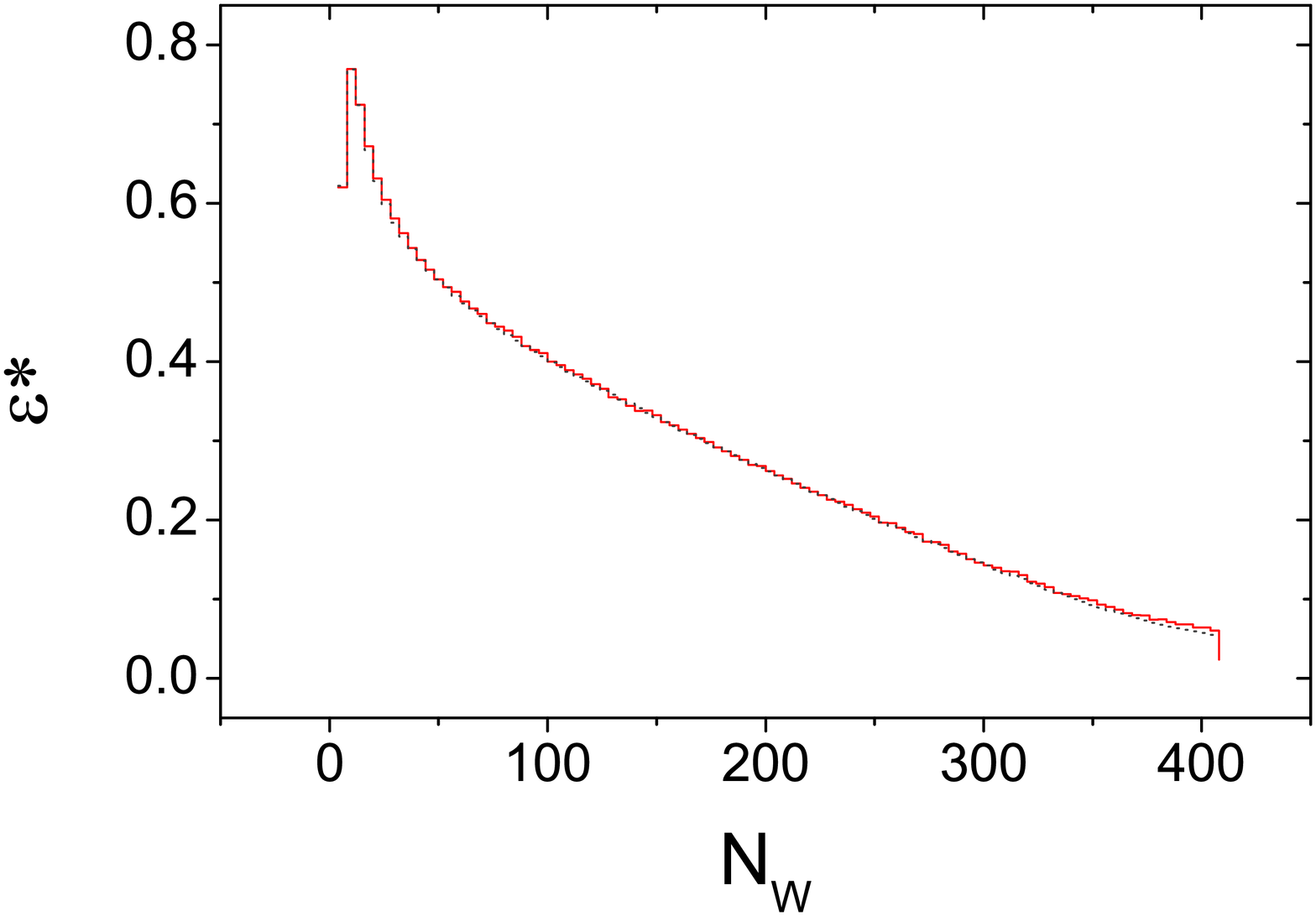}\includegraphics[angle=0,width=0.492 \textwidth]{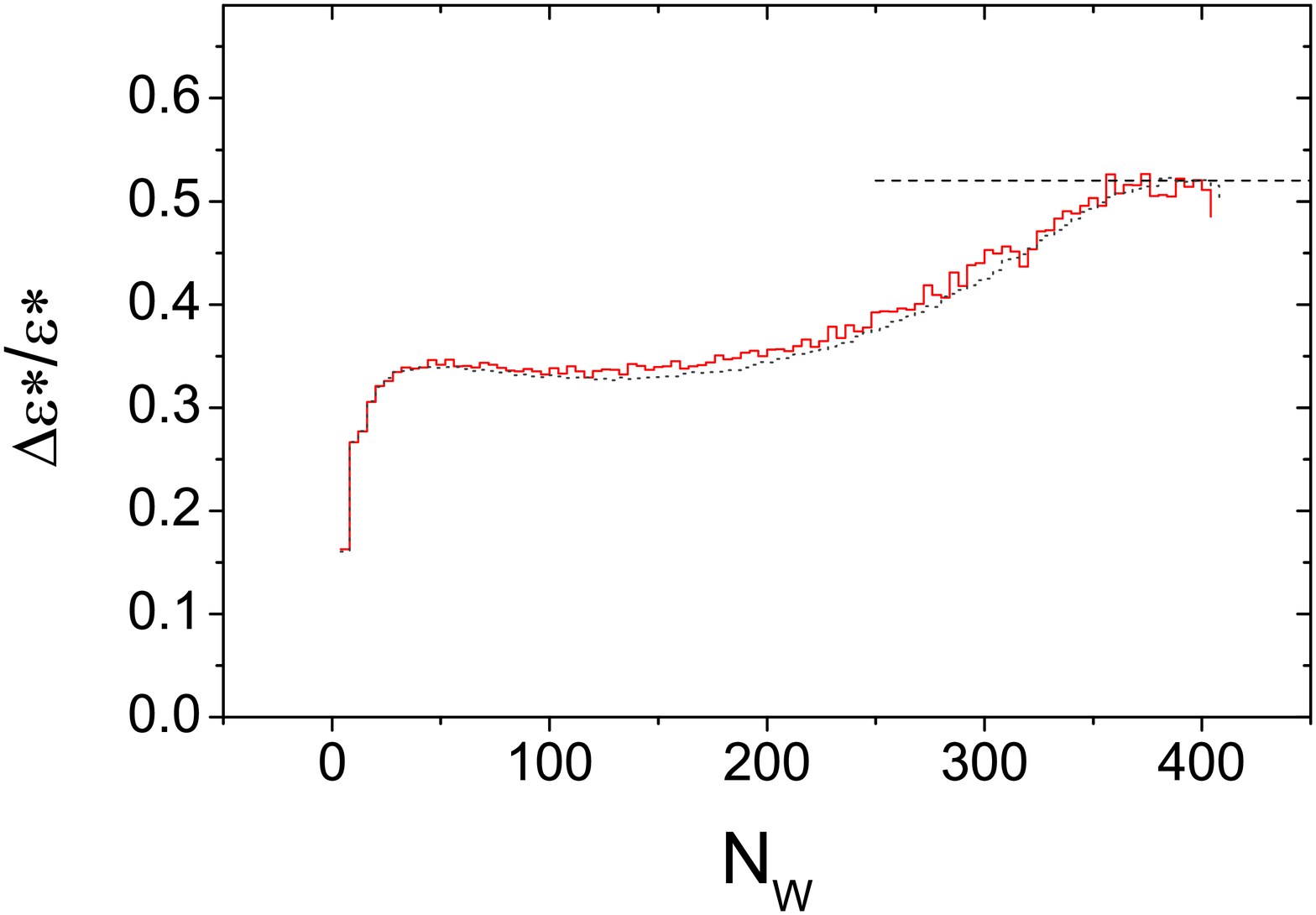}
\end{center}
\vspace{-2mm}
\caption{Various quantities computed for the ${}^{208}{\rm Pb}-{}^{208}{\rm Pb}$ collisions 
without (solid line) and with the NN correlations (dashed line) in the wounded-nucleon model for $\sigma_{\rm NN}=32~{\rm mb}$. 
The top panels show the mean total (projectile+target) number of the wounded-nucleon pairs and its scaled variance as a function of the number of the 
wounded nucleons in the projectile. The middle and bottom panels show the fixed-axes 
(standard) and variable-axes (participant) eccentricities \cite{Broniowski:2007ft} 
and their scaled standard deviation as functions of the total number of wounded nucleons. 
The horizontal line in the bottom-right panel indicates the limit $\sqrt{4/\pi-1} \simeq 0.52$ derived in \cite{Broniowski:2007ft}. 
\label{fig:var}} 
\end{figure}

Our results for the ${}^{208}{\rm Pb}-{}^{208}{\rm Pb}$ collisions are shown in Fig.~\ref{fig:var}. The notation is as follows: 
$N_W$ - the total (projectile+target) number of wounded nucleons,
$N_{\rm P}^{\rm PROJ}$ - the number of wounded nucleons in the projectile, and $N=N_W/2$ denotes the number of the wounded-nucleon pairs.  
We note that 
except for the fluctuations of the total number of the wounded-nucleon pairs plotted as a function of the wounded nucleons in the projectile
(the NA49 setup \cite{Alt:2006jr}, where the VETO calorimeter essentially measures $N_{\rm P}^{\rm PROJ}$), 
other investigated quantities are not affected by the inclusion of the NN correlations.  
While for the one-body observables from the left-side panels this is expected, as two-body correlations by 
definition do not affect one-body observables, the weakness of the effect in the fluctuation of the 
eccentricities $\varepsilon$ and $\varepsilon^\ast$ \cite{Broniowski:2007ft} is somewhat surprising.
Our conclusions for $\varepsilon$ agree with similar observations drawn in \cite{Tavares:2007mu}.

We conclude that apart for the 
multiplicity fluctuations, the neglect of the NN correlations in numerous previous studies of relativistic 
heavy-ion collisions was innocuous. In particular, the 
previous studies of $v_2$ and its fluctuations \cite{Miller:2003kd,Bhalerao:2005mm,Andrade:2006yh,Voloshin:2006gz,Alver:2006pn,%
Alver:2006wh,Sorensen:2006nw,Alver:2007rm} are not affected by the NN correlations in the nucleon distributions.

\section{$p_T$ fluctuations}

The $p_T$ fluctuations in relativistic heavy-ion collisions have been a subject of intense studies
\cite{Gazdzicki:1992ri,Stodolsky:1995ds,Shuryak:1997yj,Mrowczynski:1997kz,%
Voloshin:1999yf,Baym:1999up,Appelshauser:1999ft,Voloshin:2001ei,Prindle:2006zz,Mrowczynski:2009wk,%
Adams:2003uw,Adamova:2003pz,Adler:2003xq,Adams:2005ka,Grebieszkow:2007xz,na49:2008vb}.
Despite numerous theoretical efforts, up to now the magnitude and centrality dependence 
of these correlations has not been convincingly understood. 
In this part of the talk we present a very simple mechanism, capable of describing very well the data. 
The approach is described in a greater detail in \cite{Broniowski:2009fm}.
It generates the event-by-event 
$p_T$-fluctuations based 
on the fluctuations of the initial size of the formed system and its subsequent hydrodynamic evolution followed by
statistical hadronization (we use the single-freeze-out variant from \cite{Florkowski:2001fp,Broniowski:2001we,Broniowski:2001uk,Torrieri:2004zz}). 
Due to its statistical nature, the Glauber approach leads to an initial configuration of
the wounded nucleons (or binary collisions) which are randomly distributed. This promptly yields the fluctuations of the initial
transverse size. In short, this is the scheme: {\em smaller initial size has more compression, leading 
to faster hydrodynamic expansion, larger flow at freeze-out, and, finally, larger
transverse momenta}, and vice versa. 
We note that the effects of inhomogeneities in the initial condition 
for some observables have been studied in \cite{Hama:2009pk}.
The event-by-event fluctuations of the initial shape have been studied in detail for its elliptic
component, where they cause enhancement of the elliptic flow \cite{Aguiar:2000hw,Miller:2003kd,Bhalerao:2005mm,Andrade:2006yh,Voloshin:2006gz,Alver:2006pn,Broniowski:2007ft,Hama:2007dq,Voloshin:2007pc,%
Manly:2005zy,Alver:2006wh}. 



We define the average transverse size (in the
wounded nucleon model for the simplicity of notation) as
\begin{eqnarray}
\langle r \rangle = \sum_{i=1}^{N_W} \sqrt{x_i^2+y_i^2}, \label{eq:def}
\end{eqnarray}  
where $x_i$ and $y_i$ are coordinates of a wounded nucleon in the transverse plane. 
The original positions of nucleons in each 
nucleus are randomly generated from an appropriate Woods-Saxon distribution.
The notation $\langle \langle . \rangle \rangle$ indicates averaging over the events. 
In order to focus on the relative size of the 
effect we use the scaled standard deviation, defined for a fixed value of $N_W$ 
as $\sigma(\langle r \rangle)/\langle \langle r \rangle \rangle$. 

\begin{figure}[tb]
\begin{center}
\includegraphics[angle=0,width=0.6 \textwidth]{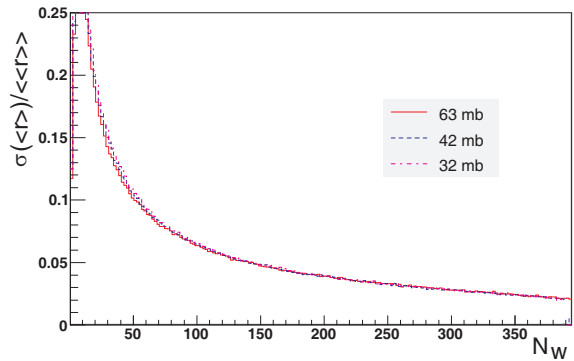} 
\end{center}
\vspace{-5mm}
\caption{Event-by-event scaled standard deviation of the size parameter $\langle r \rangle$, evaluated at fixed values 
of the number of wounded nucleons, $N_W$, for several values of $\sigma_{\rm NN}$ in  ${}^{197}{\rm Au}-{}^{197}{\rm Au}$ collisions.
\label{fig:basic}} 
\end{figure}

The results of our Monte Carlo simulations for ${}^{197}{\rm Au}-{}^{197}{\rm Au}$
performed with {\tt GLISSANDO} \cite{Broniowski:2007nz} in the wounded nucleon model are shown 
in Fig.~\ref{fig:basic}. The three curves overlap, showing insensitivity
to the value of $\sigma_{\rm NN}$ in the considered range. 
We note that the scaled standard deviation of $\langle r \rangle$ is 
about 2-3\% for central collisions, and grows towards the peripheral collisions approximately as $1/\sqrt{N_W}$.

Qualitatively very similar results are obtained for other variants, in particular for the mixed model and models with 
superimposed distribution of particles produced by each wounded nucleon \cite{Broniowski:2007nz}. We have also 
checked that using a Gaussian wounding profile $\sigma_{\rm NN}(b)$ \cite{Bialas:2006kw} for the 
NN collision, rather than the 
sharp wounding distance criterion applied here, leads to indistinguishable curves. Also,   
the use of the nucleon distributions including realistically the NN correlations, as described in the first part of this talk, 
leads to practically no difference. In other words, the behavior displayed in Fig.~\ref{fig:basic}
is robust, reflecting the random nature of the Glauber approach.

The next step, crucial in converting the size fluctuation into the observable momentum fluctuations, is hydrodynamics. 
We use the hydrodynamic approach of \cite{Broniowski:2008vp}, followed with statistical hadronization as implemented in 
{\tt THERMINATOR} \cite{Kisiel:2005hn}. 
The goal is to find how exactly the size fluctuations get converted into the $p_T$ fluctuations. 
Rather than doing tedious event-by-event hydrodynamic calculations, 
it is enough to see how much the results change when the size of the initial 
profile is scaled. The procedure presented below works, since the studied fluctuations 
follow from the initial conditions, while the differential equations of hydrodynamics are deterministic. 

The event-by-event distribution of $\langle r \rangle$ is approximately Gaussian, 
\begin{eqnarray}
f(\langle r \rangle) \sim \exp \left (-\frac{(\langle r \rangle- \langle \langle r \rangle \rangle)^2}{2 \sigma^2(\langle r \rangle)} \right ).
\label{fr}
\end{eqnarray} 
Suppose first that we run the simulations at a fixed value of $\langle r \rangle$
and as the result obtain a certain average transverse momentum, $\bar p_T$.
Because of the deterministic nature of hydrodynamics, $\bar p_T$ is a (very complicated) function of $\langle r \rangle$. 
Its value fluctuates because of the fluctuations of the initial size.  
Now, we can expand near the central value:
\begin{eqnarray}
\bar p_T - \langle \langle p_T \rangle \rangle = 
\left . \frac{d \bar p_T}{d\langle r \rangle} \right |_{\langle r \rangle = \langle \langle r \rangle \rangle} 
\left ( \langle r \rangle - \langle \langle r \rangle \rangle\right ) +\dots \label{fexp}
\end{eqnarray} 
Substituting (\ref{fexp}) into (\ref{fr}) and comparing to (\ref{fp}) we get the key formula
\begin{eqnarray}
\sigma_{\rm dyn}^2 = \sigma^2(\langle r \rangle) 
\left ( \left . \frac{d \bar p_T}{d\langle r \rangle} \right |_{\langle r \rangle = \langle \langle r \rangle \rangle} \right )^2,
\label{sigsig}
\end{eqnarray}
or, for the scaled standard deviation,
\begin{eqnarray}
\frac{\sigma_{\rm dyn}}{\langle \langle p_T \rangle \rangle} = -\frac{\sigma(\langle r \rangle)}{\langle \langle r \rangle \rangle}
\frac{\langle \langle r \rangle \rangle}{\langle \langle p_T \rangle \rangle} 
\left . \frac{d \bar p_T}{d\langle r \rangle} \right |_{\langle r \rangle = \langle \langle r \rangle \rangle}.
\label{sigsig2}
\end{eqnarray}

The full statistical distribution 
$f(\langle p_T \rangle)$  is a folding of the statistical distribution of $\langle p_T \rangle$ at a fixed 
initial size, centered around $\bar p_T$, with the distribution of $\bar p_T$ centered around $\langle\langle p_T \rangle\rangle$. 
In the central 
regions both are close to Gaussian distributions, hence we have, to a very good approximation,
\begin{eqnarray}
f(\langle p_T \rangle) \sim \int d^2{\bar{p}_T} \label{fp} 
\exp{\left ( -\frac{(\langle p_T \rangle - {\bar p}_T)^2}{2 \sigma_{\rm stat}^2} \right )}
\exp{\left ( -\frac{({\bar p}_T  - \langle \langle p_T \rangle \rangle)^2}{2 \sigma_{\rm dyn}^2}\right ) }. 
\end{eqnarray}
Carrying out the $\bar{p}_T$ integration yields the Gaussian event-by-event
distribution of $\langle p_T \rangle$ centered around $\langle \langle p_T \rangle \rangle$
with the width parameter satisfying  $\sigma^2=\sigma_{\rm stat}^2+ \sigma_{\rm dyn}^2$. 
Statistical procedures used in experimental analyses of fluctuations are designed in such a way that the dynamical component 
of the fluctuations is extracted. In our case, where the source of fluctuations is in the initial condition, we find the simple 
formula~(\ref{sigsig}), which may be compared to the experimental $\sigma_{\rm dyn}$.

The derivative in Eqs.~(\ref{sigsig},\ref{sigsig2}) 
can be computed numerically without difficulty by running just two simulations at each centrality. We perform calculations for 
initial profiles which are squeezed or stretched by 5\%.
In addition to squeezing or stretching, 
we also simultaneously adjust the central temperature in such a way, that the energy contained in the
profile is preserved. This is natural, as the total energy deposited in the transverse plane should be the same
(up to possible additional fluctuations neglected here)  for a given 
number of elementary collisions. 
Thus, we include in some sense also the temperature fluctuations discussed, {\em e.g.}, in \cite{Korus:2001au}.

\begin{figure}[tb]
\begin{center}
\includegraphics[angle=0,width=0.65 \textwidth]{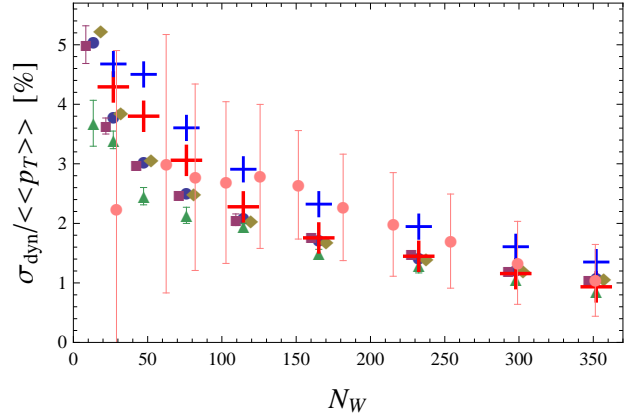} 
\end{center}
\vspace{-4mm}
\caption{Scaled dynamical transverse-momentum fluctuations, 
$\sigma_{\rm dyn}/\langle \langle p_T \rangle \rangle$ (for $\sigma_{\rm NN}=42$~mb 
in ${}^{197}{\rm Au}-{}^{197}{\rm Au}$ collisions) compared to the experimental data from STAR
\cite{Adams:2005ka} and PHENIX \cite{Adler:2003xq}. 
The lower (upper) crosses indicate our results for the wounded nucleon model (mixed model). The STAR experimental data
range from $\sqrt{s_{\rm NN}}=20$~GeV (triangles), through 130~GeV (squares), 62~GeV (diamonds), to $200$~GeV (dots). 
The PHENIX data (dots with large systematic error bars) are for 200~GeV. \label{fig:data}} 
\end{figure}  
  
Our final result is shown in Fig.~\ref{fig:data}, where we compare the model points to the
data from the STAR \cite{Adams:2005ka} and PHENIX \cite{Adler:2003xq} collaborations. The experimental cuts of 
the STAR detector have been used in our model simulations, $0.2~{\rm GeV} < p_T < 2~{\rm GeV}$. 
Looking at Fig.~\ref{fig:data}, we note a strikingly good agreement between our calculation and the experiment, in particular for the 
wounded nucleon model. The mixed model, admittedly more realistic than the 
wounded-nucleon model, overshoots the data by about 20\%. This suggests 
that the coefficient  in (\ref{sigsig2}) is somewhat too large. Its value incorporates all the dynamics 
of the system (the choice of the initial profile, details of hydrodynamics, the statistical 
hadronization), hence modifying any of these components, {\em e.g.} including viscosity effects, will lead to 
changes. We note a proper dependence on centrality, with an approximate dependence 
$\sigma_{\rm dyn}(\langle p_T \rangle)/\langle \langle p_T \rangle \rangle \sim{1}/\sqrt{N_W}$. Since the 
results of Fig.~\ref{fig:basic} very weakly depend on $\sigma_{\rm NN}$, 
to the extent that the hydrodynamic ``pushing'' is similar at various energies, our results should also 
weakly depend on the incident energy, which is a desired experimental feature in view of the STAR results.

\begin{figure}[tb]
\begin{center}
\includegraphics[angle=0,width=0.6 \textwidth]{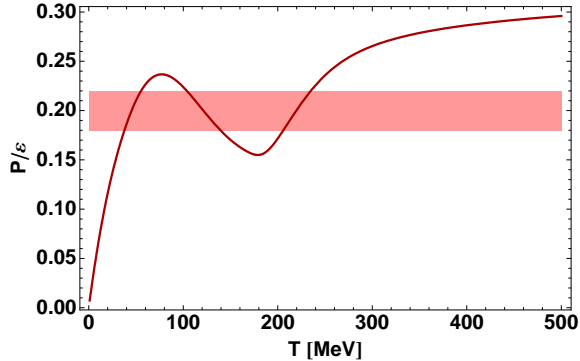} 
\end{center}
\vspace{-4mm}
\caption{Estimated {\em average} value of $P/\epsilon$ based on Eq.~(\ref{est}) (band), compared to 
the curve from the equation of state from \cite{Chojnacki:2007jc,Chojnacki:2007rq}. \label{fig:pe}} 
\end{figure}

Finally, we note that the result (\ref{sigsig2}) bears similarity to the formula derived by Ollitrault \cite {Ollitrault:1991xx}, 
where 
\begin{eqnarray}
\frac{\sigma_{\rm dyn}}{\langle \langle p_T \rangle \rangle}
 = \frac{P}{\epsilon} \frac{\sigma(\langle s\rangle)}{\langle \langle s \rangle \rangle}
 = \frac{2P}{\epsilon} \frac{\sigma(\langle r \rangle)}{\langle \langle r \rangle \rangle}, \label{olli}
\end{eqnarray}
with $s$ denoting the entropy density, $P$ - the pressure, and $\epsilon$ - the energy density. The 
second equality follows from the assumption that the total entropy deposited in the transverse plane 
depends only on the number of collisions and not on the size, hence $\langle s \rangle \sim 1/\langle r \rangle^2$ \cite{olli_priv}.
The coefficient $P/\epsilon$ in Eq.~(\ref{olli}) 
is to be understood in the averaged sense over space and time. Numerically, we find
\begin{eqnarray}
{\langle \langle r \rangle \rangle}/{\langle \langle p_T \rangle \rangle} 
\left . {d \bar p_T}/{d\langle r \rangle} \right |_{\langle r \rangle = \langle \langle r \rangle \rangle} \sim -0.4,   \label{est}
\end{eqnarray}
independent of centrality.
This yields the average $P/\epsilon$ about 0.2, which is in the expected ball park 
of realistic equations of state \cite{Chojnacki:2007jc,Chojnacki:2007rq}, see Fig.~\ref{fig:pe}%
\footnote{Presented by M. Chojnacki at V Workshop on Particle Correlations and Femtoscopy, CERN, 14-17 October 2009.}. 

\section{Conclusion}

Here are our main conclusions: 

\begin{enumerate}

\item The inclusion of the two-body NN correlations is important for the fluctuations of the 
number of produced particles. Other observables, such as the shape eccentricity fluctuations, carried over to
the $v_2$ fluctuations, are 
not affected. 

\item We reproduce the dynamical event-by-event $p_T$ fluctuations, as measured at 
RHIC by STAR and PHENIX, with the simple mechanism based on fluctuations of the initial size, which are then carried over by hydrodynamics to the 
fluctuations of the transverse flow velocity, and consequently to the transverse momentum of the produced particles. 
The hydrodynamic ``push'' is crucial in this scheme.
Other possible sources 
of fluctuations, such as the formation of 
clusters at freeze-out \cite{Broniowski:2005ae,Broniowski:2006zz,Tomasik:2008fq,Torrieri:2008zz}, 
minijets \cite{Adler:2003xq,Liu:2003jf}, or 
correlations originating from the elementary NN collisions in the corona in the core-corona 
picture \cite{Bozek:2005eu,Werner:2007bf,Bozek:2008zw,Becattini:2008ya}, 
should all be considered at the ``background'' of the fluctuations described in this talk. 

\item The hydrodynamic push is related to the equation of state, in particular to the ratio $P/\epsilon$ averaged over space and time. 
Thus, interestingly, the 
$p_T$ correlations carry, via their relation to the size fluctuations, information on the stiffness of the medium.

\end{enumerate}

One of us (WB) is grateful to Jeff Mitchell for a discussion on the PHENIX data and the conversion formulas for various measures of fluctuations, 
and to Jean-Yves Ollitrault for communicating to us the result of Eq.~(\ref{olli}).


\end{document}